\documentstyle[dina4,epsf,twocolumn]{article}
\begin{document}
\renewcommand{\textfraction}{0.1}
\renewcommand{\topfraction}{0.8}
\rule[-8mm]{0mm}{8mm}
\begin{minipage}[t]{16cm}
{\large \bf Polaronic effects in strongly coupled electron-phonon
systems: \\ Exact diagonalization results for the 2D Holstein t--J
model\\[4mm]}
H.~Fehske, G. Wellein, B. B\"auml, and H. B\"uttner\\[3mm]
{Physikalisches Institut, Universt\"at Bayreuth, 
D--95440 Bayreuth, Germany}\\[4.5mm]
\hspace*{0.5cm}
Ground--state and dynamical properties of the 2D
Holstein t--J model are examined by means of direct Lanczos diagonalization, 
using a truncation method of the phononic Hilbert space. 
The single--hole spectral function shows the formation 
of a narrow hole--polaron band as the electron--phonon coupling
increases, where the polaronic band collapse is favoured by strong Coulomb
correlations. In the two--hole sector, the hole--hole correlations 
unambiguously indicate the existence of inter--site 
bipolaronic states. At quarter--filling, a polaronic superlattice 
is formed in the adiabatic strong--coupling regime.
\end{minipage}\\[4.5mm]
\normalsize
Polaronic features of dopant--induced charge carriers have been
detected in the copper--based high-$T_c$ compounds 
$\rm La_{2-x}Sr_xCuO_{4+y}$, and even more in the
isostructural nickel--based charge--transfer oxides   
$\rm La_{2-x}Sr_xNiO_{4+y}$~\cite{BE93}. To tackle the problem 
of (bi)polaron formation in such systems exhibiting 
besides a substantial electron--phonon (EP)
coupling strong Coulomb interactions, it seems, at the moment, that 
approximation--free numerical quantum Monte--Carlo  and
exact diagonalization (ED) analyses of generic model Hamiltonians 
yield the most reliable results. Along this line, by use of ED,  
the ground--state properties of Hubbard and 
t--J models with an on--site Holstein EP coupling have been studied
on finite clusters in 1D and 2D~[2-4]. 
What is missing to date is an application of the powerful ED technique
to the calculation of {\it dynamical} properties of the Holstein t--J
model (HtJM), including the full quantum nature of phonons. 

In this contribution, we employ the Lanczos algorithm in combination
with a kernel polynomial moment expansion 
and the Maximum Entropy method~\cite{Siea96} to investigate 
the quasiparticle spectrum of a single hole--polaron
in the 2D HtJM on a ten--site square lattice. Moreover, we compute  
different hole--hole/phonon correlation functions 
at higher doping level in order to comment on hole--binding effects and  
charge-density-wave~(CDW)~formation. 

The HtJM is described by the Hamiltonian~\cite{WRF96} 
\begin{equation}
{\cal H}={\cal H}_{ph}^{}+{\cal H}^{}_{t-J}
- \sqrt{\varepsilon_p\hbar\omega}  
\sum_i \big(b_i^\dagger + b_i^{}
\big)\,\tilde{h}_i^{}\,,
\label{htjm}
\end{equation}
where ${\cal H}_{ph}$ and ${\cal H}^{}_{t-J}$ represent the phonon part 
and standard t--J model, respectively,
and the last term takes into account the 
interaction of doped holes $(\tilde{h}_i=1-\sum_\sigma
\tilde{c}_{i\sigma}^\dagger \tilde{c}_{i\sigma}^{})$ with a single
dispersionless phonon mode (which, e.g., may be thought of as
representing a local apical--oxygen coupling; 
$\varepsilon_p$ -- EP coupling constant, $\omega$ -- bare phonon
frequency). 
\begin{figure}[t]
\unitlength1mm
\begin{picture}(70,58)
\end{picture}
\end{figure}
${\cal H}$ acts in a projected Hilbert space without double
occupancy. A general state of~(\ref{htjm}) 
can be written as the direct product 
$|{\mit\Psi}\rangle= \sum_{l,k} c_l^k \; 
|l\rangle_{el} \otimes |k\rangle_{ph}$, 
where $ l$ and $k$ label the electronic and bosonic basic states,
respectively, and $ |k\rangle_{ph}=\prod_{i=1}^{N=10}
[\sqrt{n_i^k!}]^{-1} [b_i^\dagger]^{n_{i}^{k}}\,|0\rangle_{ph}$.
Since the bosonic part of the Hilbert space is infinite
dimensional we use a truncation method~\cite{WRF96} 
restricting ourselves to phononic states with at most  $M$ phonons.
To control our truncation  procedure as a function of $M$, 
we calculate the weight of the $m$--phonon states in the ground state
$|{\mit \Psi}_0\rangle$ of ${\cal H}$:     
$|c^m|^2=\sum_{l,k} |c_{l}^{k}|^2$ with $m=\sum_{i=1}^N n_i^{k}$.
In the numerical work convergence is achieved 
if the relative error of $E_0(M)$ is less than $10^{-7}$. 

\begin{figure}[t]
\centerline{\mbox{\epsfxsize 7cm\epsffile{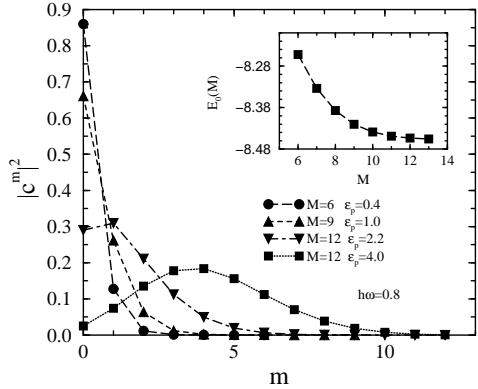}}}
\caption{Phonon--weight function  $|c^m|^2$ and 
ground--state energy $E_0(M)$ for the 2D HtJM.}
\label{F1}
\end{figure}

Figure~\ref{F1} shows $|c^m|^2$  for the 2D HtJM with a single
hole at weak, intermediate and strong EP couplings 
[in what follows we have fixed $J=0.4$ 
(all energies are measured in units of $t$)].   
The curves $|c^m|^2 (m)$ are bell-shaped and their maxima 
correspond to the most probable number of phonon quanta in the ground
state. These results, as well as the $M$--dependence of 
$E_0$ at $\varepsilon_p=4$ (see inset), confirm the importance of
multi-phonon states in the (adiabatic) strong--coupling regime
$\varepsilon_p\gg 1,\; \hbar\omega$. 

\begin{figure}[t]
\centerline{\mbox{\epsfxsize 7cm\epsffile{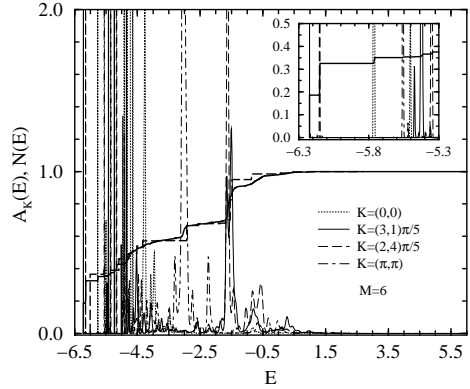}}}
\centerline{\mbox{\epsfxsize 7cm\epsffile{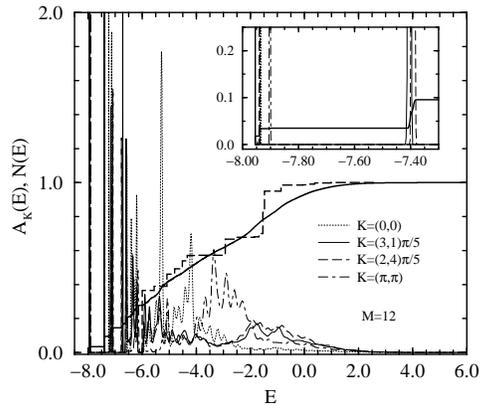}}}
\caption{Single--hole spectral function $A_{\vec{K}}(E)$ 
for the 2D HtJM at $\hbar\omega=0.8$ and $\varepsilon_p=0.5$ 
(upper panel) 3.4 (lower panel), where the energy scale is shifted by
$E_0^{(N)}$. The insets 
show the low-energy part of the spectra. The integrated density of
states  
$N(E)=\int_{-\infty}^E dE^{\prime} \sum_{\vec{K}}
A_{\vec{K}}(E^{\prime})$ 
(bold solid line) is depicted in comparison with the result for the
pure t--J model (dashed line).}
\label{F2}
\end{figure}

In the analysis of the HtJM we start with the study 
of just a single dynamic hole. Increasing the EP coupling in the
adiabatic regime, we notice a continuous but rather sharp 
crossover from nearly--free polaron, described by an 
effective transfer amplitude that is only weakly
reduced from its value in the pure t--J model, to a less mobile
(small--size) adiabatic Holstein hole--polaron (AHP). 
Moreover, we found that the critical EP coupling for the polaron transition is
substantially reduced due to prelocalization effects of the hole 
in the antiferromagnetic spin background~\cite{Feea9395}. 
To elucidate the difference between the FP and AHP limits 
and to demonstrate the formation of a hole--polaron band at large
$\varepsilon_p$, in Fig.~\ref{F2} we present 
the results for the $\vec{K}$--resolved
spectral function
\begin{eqnarray}
A_{\vec{K}}(E)^{}\!\!\!&=&\!\!\! \sum_{n,\sigma} |\langle {\mit\Psi}_n^{(N-1)}
|\,\tilde{c}_{\vec{K}\sigma}^{}\,|\,{\mit\Psi}_0^{(N)}\rangle|^2\qquad\nonumber\\
&&\quad\times\delta [E-(E_n^{(N-1)}-E_0^{(N)})]\,.
\label{spfu}
\end{eqnarray}
Of course in the very weak--coupling regime  
the spectral function is barely changed from that of the pure t--J model. 
Increasing  $\varepsilon_p$,  the lowest peaks in each
$A_{\vec{K}}$ separate from the rest of the spectrum. 
These states become very close in energy and a narrow
well--separated energy band evolves in the 
strong--coupling case, where the gap 
to the next higher ``band'' is of the order
of the phonon frequency $\hbar\omega$. 
Note that the transition to the AHP state is accompanied 
by a strong increase in the on--site hole--phonon 
correlations which are about one
order in magnitude larger than the nearest--neighbour (NN) ones
(cf. Fig. 11 in Ref.~\cite{WRF96}).  
This indicates that the AHP quasiparticle comprising a  
`quasi--localized' hole and the phonon cloud is mainly confined to a
single lattice site. As the phonon frequency is enlarged at fixed, 
the hole--phonon correlations are smeared out, and the crossover to the small
hole--polaron is shifted to larger values of the EP coupling.

\begin{figure}[t]
\centerline{\mbox{\epsfxsize 6cm\epsffile{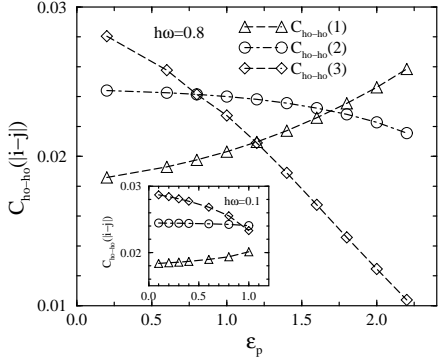}}}
\caption{Non--equivalent hole--hole pair correlation functions 
$C_{ho-ho}(|i-j|)$ in the two--hole ground state 
of the HtJM  as a function of $\varepsilon_p$; here    
1--3 label NN, next NN, and third NN distances.}
\label{F3}
\end{figure}

Next we wish to discuss  the two--hole problem. To
get a feel for hole--binding effects, we have calculated
the hole--hole correlation function  
\begin{equation}
C_{ho-ho}^{}(|i-j|)=\langle {\mit\Psi}_0(\varepsilon_p,J)
|\tilde{h}_i^{} \tilde{h}_j^{}|{\mit\Psi}_0(\varepsilon_p,J)\rangle\,.
\label{choho}
\end{equation}  
Results for $C_{ho-ho}^{}(|i-j|)$ are given in Fig.~\ref{F3}. 
In the weak--coupling region, $C_{ho-ho}(|i-j|)$
becomes maximum at the largest distance of the ten--site lattice,
while in the intermediate EP coupling regime 
the preference is on next NN pairs.
As expected, increasing further $\varepsilon_p$, 
the maximum in $C_{ho-ho}(|i-j|)$ is shifted to the 
shortest possible distance, indicating hole--hole attraction. 
At $\varepsilon_p\gg 1 $, the two holes become `self--trapped'
sharing a sizeable common lattice distortion, i.e., 
a nearly immobile hole--bipolaron is formed.  
The behaviour of $C_{ho-ho}$ is found to be qualitatively
similar for higher (lower) phonon frequencies (see inset), 
except that the crossings of different hole--hole correlation 
functions occur at larger (smaller) values of~$\varepsilon_p$,
which shows the importance of {\it both} parameter ratios $\varepsilon_p/t$
and $\sqrt{\varepsilon_p/\hbar\omega}$. 

\begin{figure}[t]
\centerline{\mbox{\epsfxsize 3.5cm\epsffile{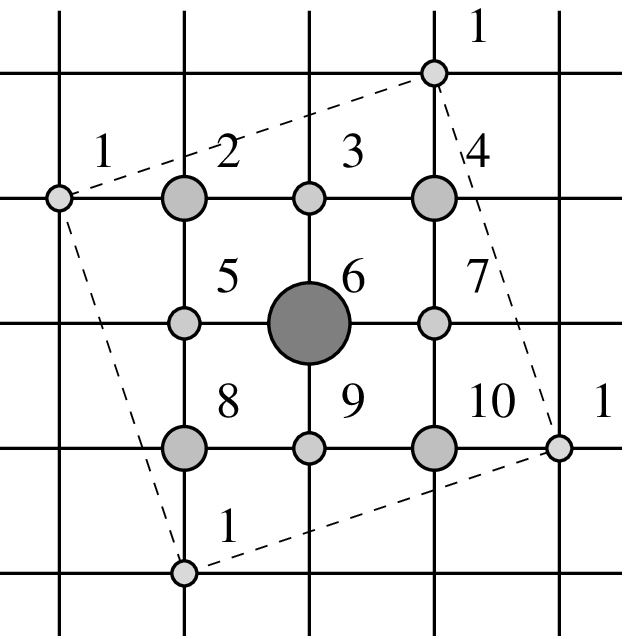}} 
\mbox{\epsfxsize 3.5cm\epsffile{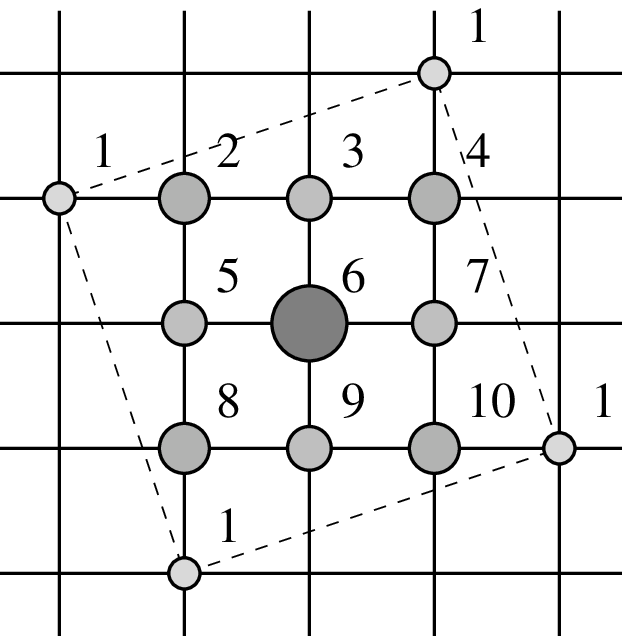}}}
\caption{$C_{ho-ho}(|6-j|)$ [left] 
and $C_{ho-ph}(|6-j|)$ [right] are displayed at $\varepsilon_p=3$ 
and $\hbar\omega=0.8$, where both diameter and gray level of the 
circles being proportional to the correlation strength.}
\label{F4}
\end{figure}

Finally let us consider the quarter--filled band case. 
Here, we have investigated the more simple spinless fermion model 
(total $S^z=S^z_{max}$). In accordance with previous approximative treatments 
based on the inhomogeneous variational Lang--Firsov
approach~\cite{Feea9395}, we found, as the EP coupling increases,
evidence for a transition from a FP state    
to a 2D polaronic superlattice, where the holes 
are self--trapped on every each other site.   
This crossover is signaled by a pronounced peak in the charge
structure factor at $(\pi,\pi)$. 
To visualize the correlations in this state in more detail, in Fig.~\ref{F4} 
we have depicted $C_{ho-ho}(|i-j|)$ and  
the corresponding hole--phonon density correlation function 
$C_{ho-ph}(|i-j|)=\langle {\mit\Psi}_0^{}|\tilde{h}_i^{} b^\dagger_j
b^{}_j|{\mit\Psi}_0^{}\rangle$ as a function of $|i-j|$.
Our exact results clearly show the phonon--dressing of the holes 
and the resulting tendency towards CDW formation.
A similar polaron ordering was observed in $\rm
La_{1.5}Sr_{0.5}NiO_4$.   
 
This work was performed under the auspices of
Deutsche For\-schungsgemeinschaft, SFB 279, Bayreuth.
We thank the LRZ (M\"unchen) and the GMD (St. Augustin) 
for allocation of CPU time on the IBM SP2 parallel computers.
We are particularly indebted to R. N. Silver for putting his Maximum
Entropy code at our disposal.
\vspace*{-0.3cm}
\bibliography{ref}
\bibliographystyle{phys}
\end{document}